\documentclass[11pt,a4paper]{article}

\usepackage{amsmath}
\usepackage{graphicx}
\usepackage{subfig}
\usepackage{cite}
\usepackage{url}
\usepackage{multirow}

\begin{document}

\title{Mathematical Modelling, Simulation, and Optimal Control
of the 2014 Ebola Outbreak in West Africa\thanks{This is a preprint
of a paper whose final and definite form is
\emph{Discrete Dynamics in Nature and Society} (Print ISSN: 1026-0226;
Online ISSN: 1607-887X) 2015, Article ID 842792, 9~pp.
See {\tt http://dx.doi.org/10.1155/2015/842792}.
Submitted 26-Dec-2014; revised 27-Feb-2015; accepted 28-Feb-2015.}}

\author{Amira Rachah$^\dag$\\
{\tt arachah@math.univ-toulouse.fr}
\and Delfim F. M. Torres$^*$\\
{\tt delfim@ua.pt}}

\date{$^\dag$Math\'{e}matiques pour l'Industrie et la Physique,\\
Institut de Math\'{e}matiques de Toulouse, Universit\'e Paul Sabatier,\\
31062 Toulouse Cedex 9, France\\[0.3cm]
$^*$Center for Research and Development in Mathematics and Applications (CIDMA),
Department of Mathematics,\\
University of Aveiro, 3810-193 Aveiro, Portugal}

\maketitle


\begin{abstract}
The Ebola virus is currently one of the most virulent pathogens for humans.
The latest major outbreak occurred in Guinea, Sierra Leone and Liberia in 2014.
With the aim of understanding the spread of infection in the affected countries,
it is crucial to modelize the virus and simulate it. In this paper, we begin by studying
a simple mathematical model that describes the 2014 Ebola outbreak in Liberia. Then,
we use numerical simulations and available data provided by the
World Health Organization to validate the obtained mathematical model.
Moreover, we develop a new mathematical model including vaccination of individuals.
We discuss different cases of vaccination in order to predict the effect
of vaccination on the infected individuals over time. Finally, we apply optimal control
to study  the impact of vaccination on the spread of the Ebola virus.
The optimal control problem is solved numerically by using
a direct multiple shooting method.
\end{abstract}


\section{Introduction}

Ebola is a lethal virus for humans. It was previously confined to Central Africa
but recently was also identified in West Africa \cite{barraya}.
As of October 8, 2014, the World Health Organization (WHO) reported 4656 cases of Ebola virus deaths,
with most cases occurring in Liberia \cite{joseph}. The extremely rapid increase
of the disease and the high mortality rate make this virus a major problem for public health.
Typically, patients present fever, malaise, abdominal, headache, and asthenia.
One week after the onset of symptoms, a rash often appears followed by haemorrhagic
complications, leading to death after an average of 10 days in 50\%--90\% of infections
\cite{anon1,anon2,okwar,legrand}. Ebola is transmitted through direct contact with blood,
bodily secretions and tissues of infected ill or dead humans and non-human primates
\cite{dowel,borio,peter}.

The main aim of this work is to understand the dynamics
of the Liberian population infected by Ebola virus in 2014,
by using an appropriate mathematical model. More precisely, we begin by considering
a system of ordinary differential equations to describe the 2014 Ebola outbreak,
which is nothing else than an epidemic SIR model, that is, a model
based on the division of the population into three groups: the Susceptible,
the Infected, and the Recovered \cite{Gerard,zeng,chow}.
We simulate the model obtained by parameters estimated on November 4, 2014,
by Kaurov \cite{kaurov}. Our objectives are to better understand
the outbreak and to predict the effect of vaccination
on the infected individuals over time.

In order to deal with the epidemic of Ebola, governments have decided to implement tough measures.
For example, some have applied quarantine procedures while others have opted for
mass vaccination plans as a precaution to the epidemic \cite{vacc1,vacc2,vacc3,delf}.
In this context, we consider an optimal control problem to study
the effect of vaccination on the spread of virus.

The text is organized as follows. In Section~\ref{sec:2} we introduce
a basic mathematical model to describe the dynamics of the Ebola virus
during the 2014 outbreak in Liberia. After the mathematical modelling,
we use the obtained model to simulate it in Section~\ref{sec:3} with
the parameters estimated from recent statistical data based
on the WHO report of the 2014 Ebola outbreak \cite{who}.
Following \cite{kaurov}, in Section~\ref{sec:vac}
we study different cases of  vaccination
of individuals by adding a vaccination term to the model.
Section~\ref{sec_control} presents an optimal control problem,
which we use to study the impact of a vaccination campaign
on the spread of the virus. We end with Section~\ref{sec:4}
of conclusions and future work, where the results are summarized
and some research perspectives presented.


\section{The Basic Mathematical Model}
\label{sec:2}

The objective of this section is to describe mathematically the
dynamics of the population infected by the Ebola virus.
The dynamics is described by a system of differential equations.
This system is based on the common SIR (Susceptible-Infectious-Recovery)
epidemic model, where the population is divided into three groups:
the susceptible group, denoted by $S(t)$, the infected group, denoted by $I(t)$,
and the recovered group, denoted by $R(t)$. The total population, assumed constant
during the short period of time under study, is given by  $N = S(t) + I(t) + R(t)$.

\begin{figure}
\centering
\includegraphics[scale=0.95]{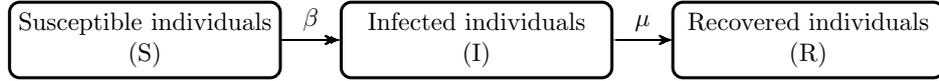}
\caption{The scheme of the Susceptible-Infected-Recovered model \eqref{eq:S}--\eqref{eq:R}.}
\label{SIR_fig1}
\end{figure}

To create the equations that describe the population of each group along time,
let us start with the fact that the population of the susceptible group will be reduced
as the infected come into contact with them with a rate of infection $\beta$.
This means that the change in the population of susceptible is equal to the negative
product of $\beta$ with $S(t)$ and $I(t)$:
\begin{equation}
\label{eq:S}
\dfrac{dS(t)}{dt} = -\beta \dfrac{S(t)I(t)}{N}.
\end{equation}
Now, let us create the equation that describes the infected group over time, knowing
that the population of this group changes in two ways:
\begin{description}
\item[(i)] people leave the susceptible group and join the infected group,
thus adding to the total population of infected a term $\beta S(t)I(t)$;

\item[(ii)] people leave the infected group and join the recovered group,
reducing the infected population by $-\mu I(t)$.
\end{description}
This is written as
\begin{equation}
\label{eq:I}
\dfrac{dI(t)}{dt} = \beta \dfrac{S(t)I(t)}{N} - \mu I(t).
\end{equation}
Finally, the equation that describes the recovery population
is based on the individuals recovered from the virus at rate $\mu$.
This means  that the recovery group is increased by $\mu$ multiplied by $I(t)$:
\begin{equation}
\label{eq:R}
\dfrac{dR(t)}{dt} = \mu I(t).
\end{equation}
Figure~\ref{SIR_fig1} shows the relationship between the three variables of our SIR model.
As shown in the next section, this simple model describes well
the 2014 Ebola outbreak in Liberia for suitable chosen
parameters $\beta$ and $\mu$ and appropriate initial
conditions $S(0)$, $I(0)$ and $R(0)$.


\section{Numerical Simulations of the Basic Model and Discussion}
\label{sec:3}

\begin{figure}
\centering
\includegraphics[width=10cm]{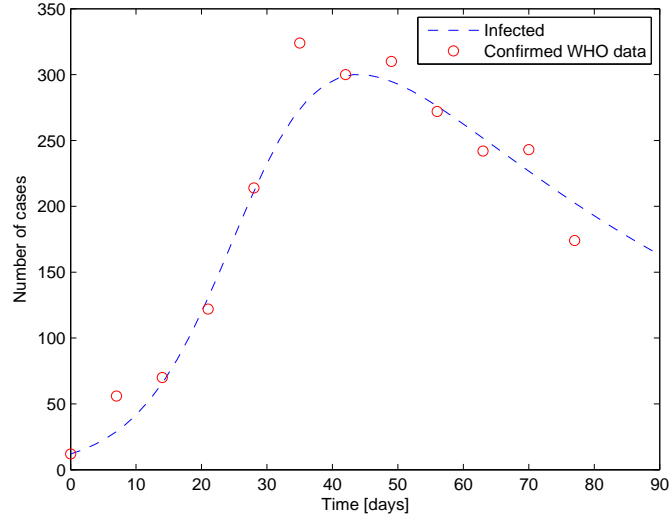}
\caption{Graph of Infected $I(t)$ obtained from \eqref{eq:SIR}
with $\beta=0.000318$, $\mu=0.0175$, $S(0)=460$, $I(0)=12$, and $R(0)=0$
\emph{versus} the real data of confirmed cases for the 2014 Ebola outbreak
occurred in Liberia from July 14 (day $t=0$) and October 12, 2014 (day $t=90$).}
\label{fit_data_who}
\end{figure}

\begin{figure}
\centering
\includegraphics[width=10cm]{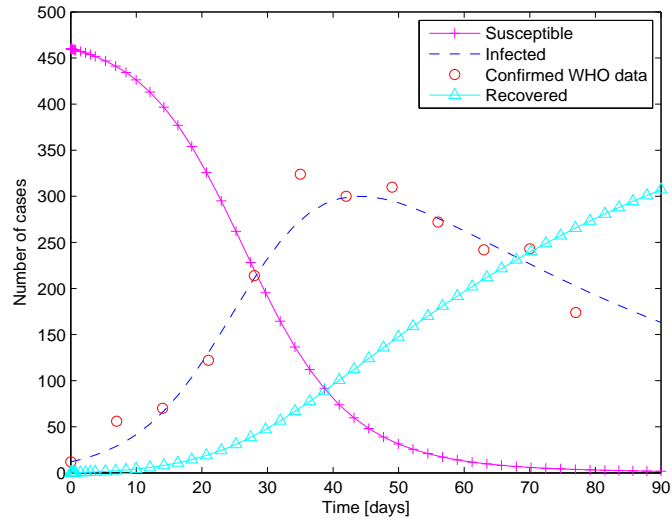}
\caption{The evolution of susceptible ($S(t)$), infected ($I(t)$), and recovered ($R(t)$)
for the period July 14 ($t=0$) and October 12, 2014 ($t=90$), given by the Ebola Liberia
model \eqref{eq:SIR} with $\beta=0.000318$, $\mu=0.0175$, $S(0)=460$, $I(0)=12$, and $R(0)=0$.
The real data of confirmed cases for the 2014 Ebola outbreak occurred in Liberia,
taken from \cite{who}, is shown with circles.}
\label{result1}
\end{figure}

With the aim to approximate the 2014 Ebola outbreak in Liberia,
we now simulate the complete set of equations \eqref{eq:S}--\eqref{eq:R}
that describe the SIR model:
\begin{equation}
\label{eq:SIR}
\begin{cases}
\dfrac{dS(t)}{dt} = -\beta \dfrac{S(t)I(t)}{N},\\[0.3cm]
\dfrac{dI(t)}{dt} = \beta \dfrac{S(t)I(t)}{N} - \mu I(t),\\[0.3cm]
\dfrac{dR(t)}{dt} = \mu I(t),
\end{cases}
\end{equation}
where $\beta$ is the rate of infection and $\mu$ the rate of recovery.
The data for the outbreak in Liberia is based on the cumulative numbers
of total reported cases (confirmed and probable) taken from the
official data reported by the Ministry of Health of Liberia
up to October 26, 2014, as provided by the World Health Organisation (WHO) \cite{who}
(in particular, the Ebola virus disease cases reported each week from Liberia
are given in Figure~2 of \cite{who:fig}).
The model fits well the reported data of confirmed (means infected) cases in Liberia.
Note that the exact population size $N$ does not need to be known to estimate
the model parameters as long as the number of cases is small compared
to the total population size \cite{althaus}. To estimate the parameters of
the model, we adapted the initialisation of $S$ and $I$ with the reported data
of \cite{who:fig} by fitting the actual data of confirmed
cases in Liberia. The data that we are studying is the number of individuals
with confirmed infection between July 14 and October 12, 2014.
The result of fitting is shown in Figure~\ref{fit_data_who}.
The comparison between the curve of infected obtained by our simulation
and the reported data of confirmed cases by WHO shows that the mathematical
model \eqref{eq:SIR} fits well the real data by using $\beta=0.000318$
as the rate of infection and $\mu=0.0175$ as the rate of recovery.
The initial numbers of susceptible, infected, and recovered groups, are given by
$S(0)=460$, $I(0)=12$, and $R(0)=0$, respectively. The choice of $S(0)$
is in agreement with the data shown in Figure~2 of \cite{who:fig},
as the biggest value in the histogram is 460. To interpret the simulation results,
we describe the evolution of each group of individuals over time and the link between them.
Figure~\ref{result1} shows that the susceptible group begins to plummet
and, simultaneously, the number of infected begins to rise. This is
due to how infectious the Ebola virus is.

When we compare our results with those of Astacio et al. \cite{astacio},
when they modeled the 1976 outbreak of Ebola in Yambuku,
a small village in Mongala District in northern Democratic Republic
of Congo (previously Zaire), and the 1995 outbreak of Ebola in Kikwit,
in the southwestern part of the Democratic Republic of Congo,
we note that the evolution of each group obtained in our simulations is reasonable.
The main difference between the 2014 outbreak of Ebola in Liberia and the previous
outbreaks of Ebola in Yambuku and Kikwit studied in \cite{astacio}
is at the end of the evolution of infected:  while in the previous
1976 and 1995 outbreaks of Ebola in Yambuku and Kikwit the infected group
goes to zero, now in the 2014 outbreak of Ebola in Liberia
the infected group remains important with a high number of infected individuals
by the virus, which describes the reality of what is currently happening in Liberia.
One can conclude that the 2014 Ebola outbreak of Liberia is much more dangerous than
the previous Ebola outbreaks.


\section{Modelling and Simulation of Ebola Outbreak with Vaccination}
\label{sec:vac}

In this section, we simulate the model by using parameters estimated on November 4, 2014,
by Kaurov \cite{kaurov}, who studied statistically the recent data
of the World Health Organisation (WHO) \cite{who}.
In his statistical study, Kaurov studied the outbreak by modeling it
with Wolfram's \textsf{Mathematica} language. Here we use his SIR model for modelling
and time-discrete equations for resolution of the set of equations through \textsf{Matlab}.


\subsection{The Basic Model with Kaurov's Estimation of Parameters}
\label{sec:Kep}

Let us recall the complete set of equations considered by \cite{kaurov}:
\begin{equation}
\label{SIR_model}
\begin{cases}
\dfrac{dS(t)}{dt} = -\beta S(t)I(t),\\[0.3cm]
\dfrac{dI(t)}{dt} = \beta S(t)I(t) - \mu I(t),\\[0.3cm]
\dfrac{dR(t)}{dt} = \mu I(t),
\end{cases}
\end{equation}
where, as before, $\beta$ is the rate of infection and $\mu$ is the rate of recovery.
The parameters obtained by Kaurov, knowing that 95\%
of population is susceptible and 5\% of population is infected,
are $\beta=0.2$ and $\mu=0.1$ \cite{kaurov}.
In agreement, the initial susceptible, infected, and recovered populations
are given, respectively, by $S(0)=0.95$, $I(0)=0.05$, and $R(0)=0$
(note that in contrast with Sections~\ref{sec:2} and \ref{sec:3},
the variables are now dimensionless).

\begin{figure}
\centering
\includegraphics[width=10cm]{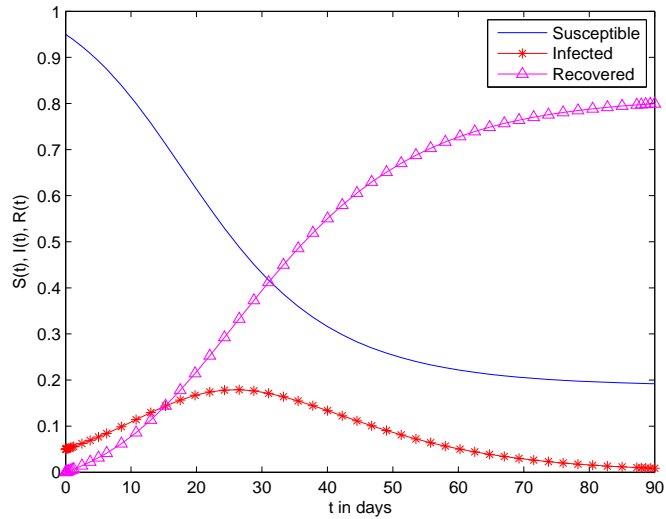}
\caption{Solution to the 2014 outbreak Ebola model \eqref{SIR_model}
with $\beta=0.2$, $\mu=0.1$, $S(0)=0.95$, $I(0)=0.05$, and $R(0)=0$.\label{result_SIR1}}
\end{figure}

Our numerical simulations were done with the ODE solver of \textsf{Matlab}.
The results are shown in Figure~\ref{result_SIR1}. To interpret the results,
we describe the evolution of each group over time and the links between them.
Figure~\ref{result_SIR1} shows that the susceptible group begins to plummet
due to how infectious the virus is and, simultaneously, the infected
group's number begins to rise.

An important dimensionless quantity that plays a key role in any SIR model
is the well-known basic reproduction number $R_{0}$ \cite{reproduct}.
This number tells us how fast the disease will spread at the epidemic.
For model \eqref{SIR_model}, the basic reproduction number is simply given by $\beta/\mu$.
Let us now discuss the comparison between the different values of $R_0$ obtained recently.
On September 2, 2014, Althaus studied the case of the 2014 Ebola outbreak occurred in Guinea,
in Sierra Leone, and in Liberia \cite{althaus}. In his study, Althaus used the WHO data occurred
in the three countries. He found different values for the basic reproduction number,
depending on the country: in Guinea $R_0 = 1.51$, in  Sierra Leone the value of $R_0$ is given by $2.53$,
while in Liberia $R_0$ is equal to $1.59$. Let us now compare between these values and the value
for the basic reproduction number studied by Kaurov, knowing that his study, which was done on November 4, 2014,
is more recent than the one of Althaus, done September 2, 2014. As the parameters estimated statistically
by Kaurov are for the three countries, our comparison will be between $R_{0,A}$, the mean of the three
basic reproduction numbers found by Althaus, and $R_{0,K}$, the number found more recently by Kaurov.
The value of $R_{0,A}$ is given by $1.876$, where the more recent $R_{0,K}$ is given by $2$.
We claim that the basic reproduction number increased
between September and November 2014.


\subsection{Dynamics in Case of Vaccination}

Infectious diseases have tremendous influence on human life.
In last decades, controlling infectious diseases has been an increasingly complex issue.
A strategy to control infectious diseases is through vaccination. Now our idea is to study the
effect of vaccination in practical Ebola situations. The case of the French nurse cured
of Ebola with the help of an experimental vacine is a proof of the possibility of treatment
by vaccinating infected individuals \cite{valler}.

\begin{center}
\begin{figure}
\includegraphics[scale=0.95]{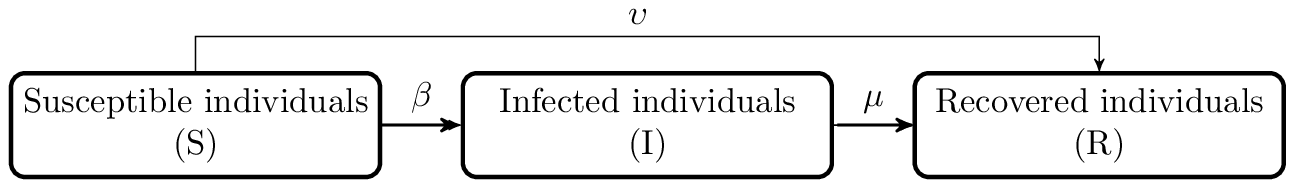}
\caption{The scheme of the Susceptible-Infected-Recovered model
with vaccination \eqref{SIR_vac}.\label{SIR_fig_vac}}
\end{figure}
\end{center}

In this section, we improve the SIR model described in Section~\ref{sec:Kep}.
The system of equations that describe the Susceptible-Infectious-Recovery (SIR)
model with vaccination is:
\begin{equation}
\label{SIR_vac}
\begin{cases}
\dfrac{dS(t)}{dt} = -\beta S(t)I(t) - \upsilon S(t),\\[0.3cm]
\dfrac{dI(t)}{dt} = \beta S(t)I(t) - \mu I(t),\\[0.3cm]
\dfrac{dR(t)}{dt} = \mu I(t) + \upsilon S(t),
\end{cases}
\end{equation}
where $\beta$ is the rate of infection, $\mu$ is the rate of recovery,
and $\upsilon$ is the percentage of individuals vaccinated every day \cite{insa}.
Figure~\ref{SIR_fig_vac} shows the relationship between
the variables of the SIR model with vaccination \eqref{SIR_vac}.

\subsection{Numerical Simulation in Case of Vaccination}

We simulate the SIR model with vaccination \eqref{SIR_vac}
in order to predict the evolution of every group of individuals
in case of vaccination. Our study is based in the test of different
rates of vaccinations and their effect on the curve of every group.
\begin{figure}
\centering
\subfloat[$\upsilon=0$]
{\includegraphics[scale=0.48]{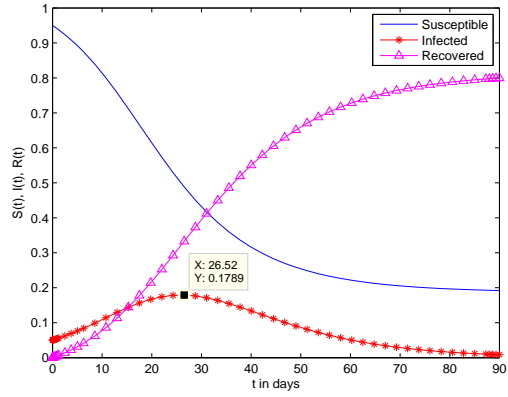}}\\
\subfloat[$\upsilon=0.005$]
{\includegraphics[scale=0.48]{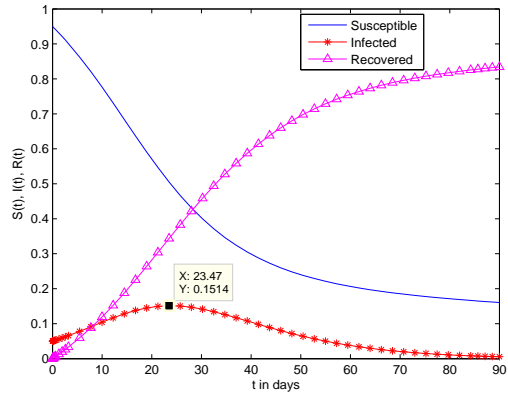}}\\
\subfloat[$\upsilon=0.01$]
{\includegraphics[scale=0.48]{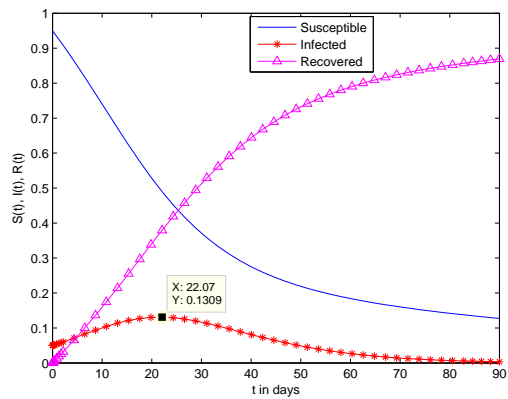}}
\caption{A solution to the SIR model \eqref{SIR_vac}
with $\beta=0.2$, $\mu=0.1$, $S(0)=0.95$, $I(0)=0.05$, $R(0)=0$,
and different rates of vaccination: $\upsilon=0$ (a), $\upsilon=0.005$ (b),
and $\upsilon=0.01$ (c).\label{SIR_vac_simul1}}
\end{figure}
\begin{figure}
\centering
\subfloat[$\upsilon=0.02$]
{\includegraphics[scale=0.48]{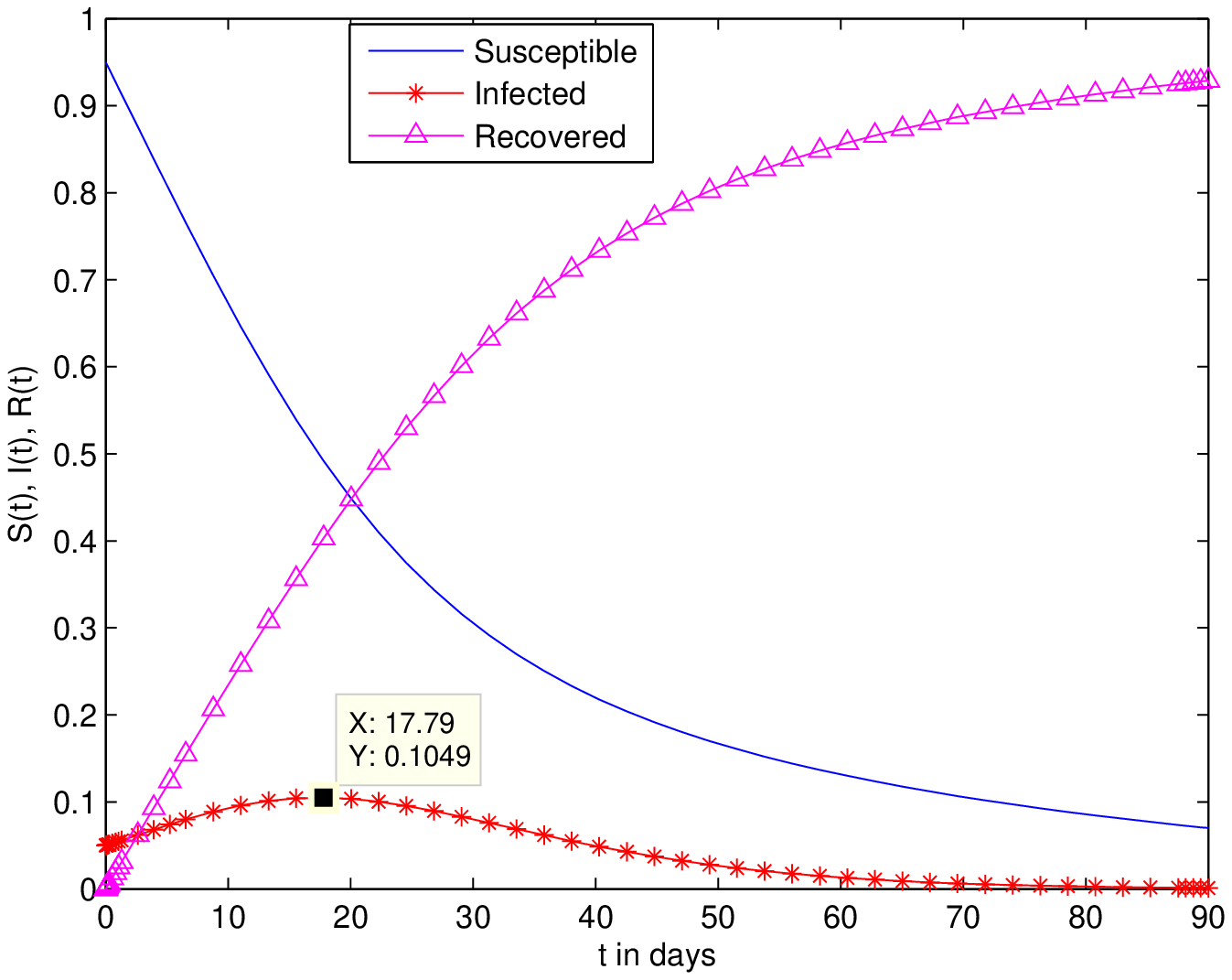}}\\
\subfloat[$\upsilon=0.03$]
{\includegraphics[scale=0.48]{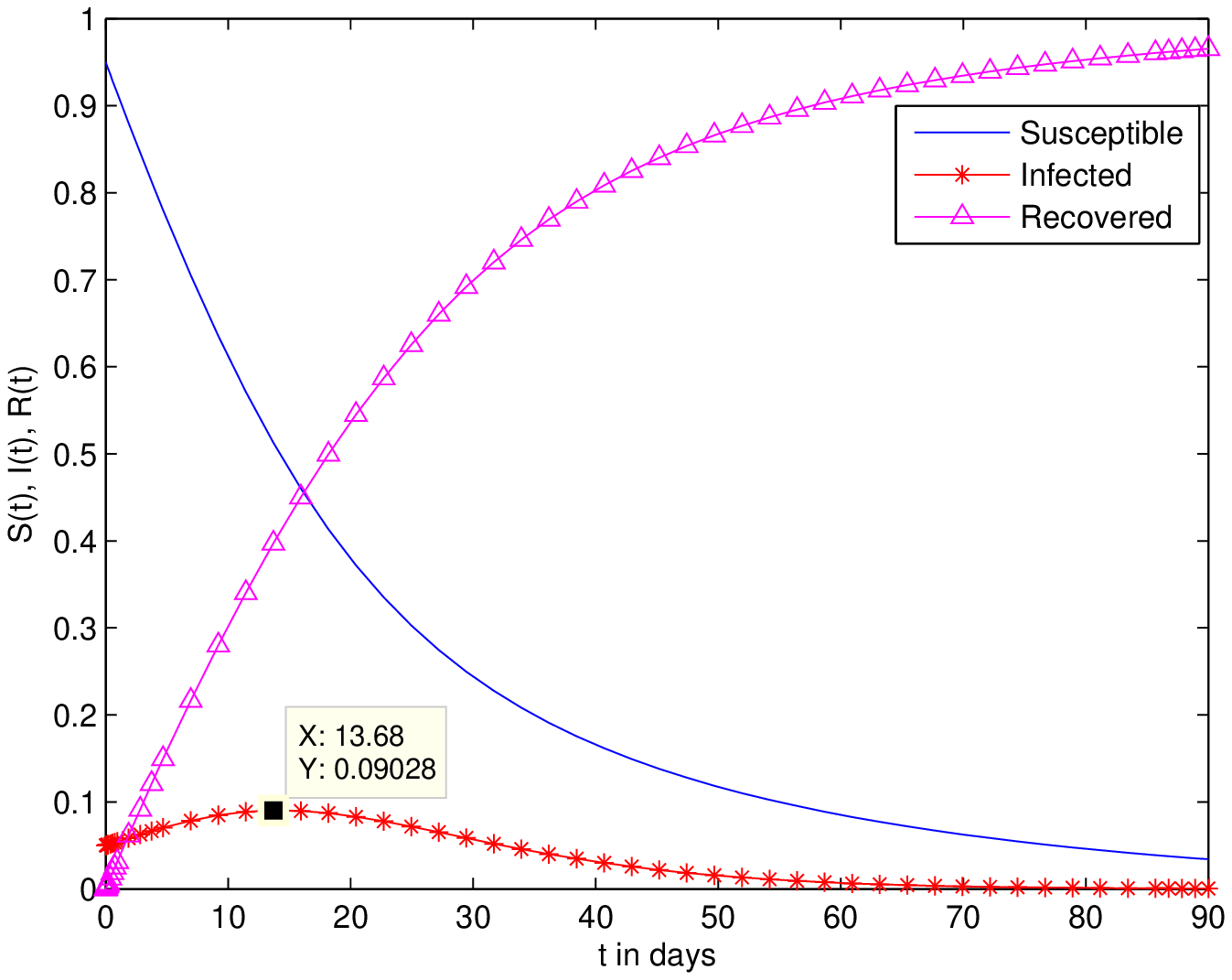}}\\
\subfloat[$\upsilon=0.06$]
{\includegraphics[scale=0.48]{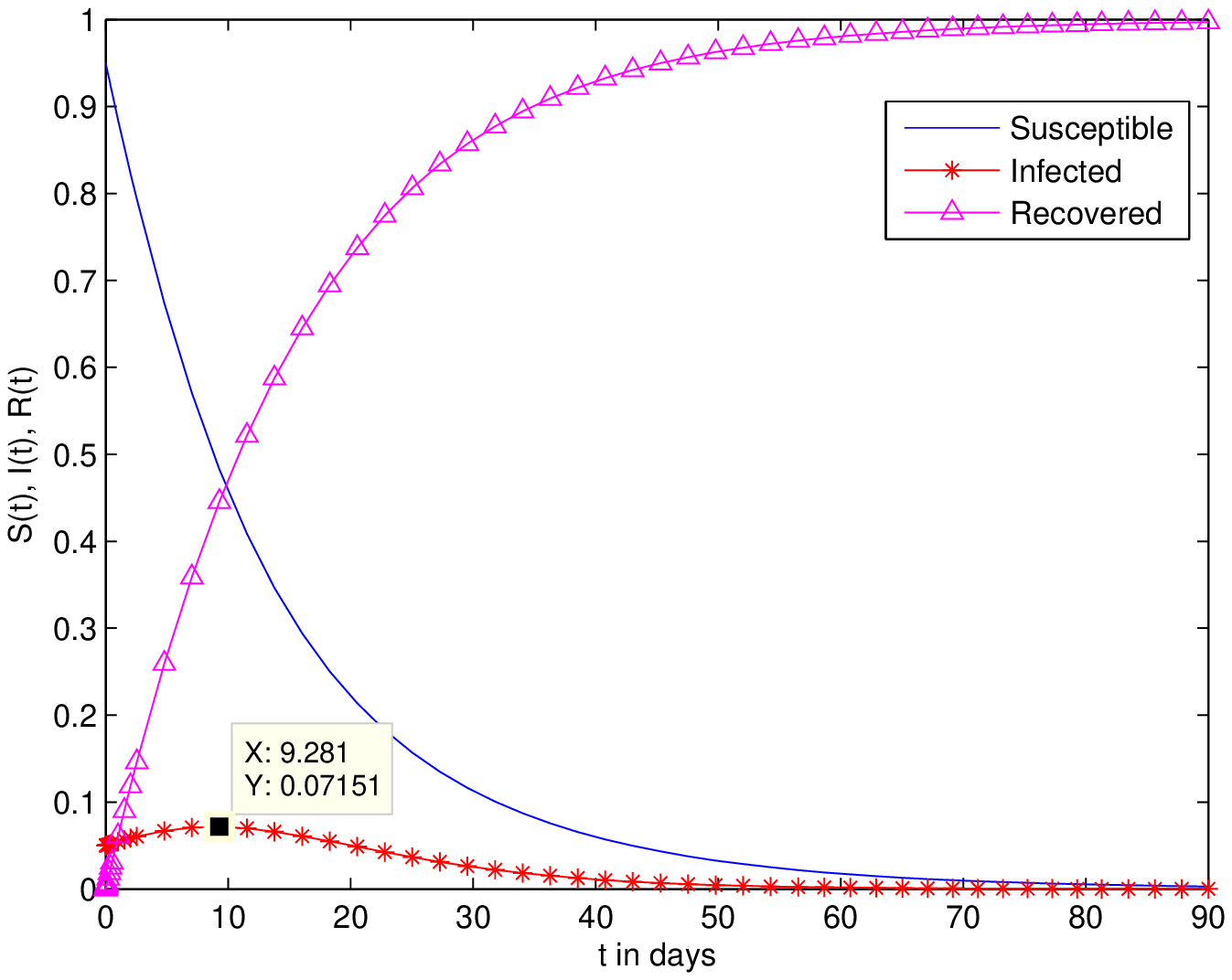}}
\caption{A solution to the SIR model \eqref{SIR_vac}
with $\beta=0.2$, $\mu=0.1$, $S(0)=0.95$, $I(0)=0.05$, $R(0)=0$,
and different rates of vaccination: $\upsilon=0.02$ (a), $\upsilon=0.03$ (b),
and $\upsilon=0.06$ (c).\label{SIR_vac_simul2}}
\end{figure}
The results are shown in Figures~\ref{SIR_vac_simul1} and \ref{SIR_vac_simul2}.
These figures present different rates of vaccination and their effect on the
peak of the curve of infected individuals along time. The time-dependent
curve of infected individuals shows that when the percentage of vaccinated individuals increases,
the peak of the curve of infected individuals is less important and the period of infection
(the corresponding number of days) is shorter while the number of recovered gets
more important. In fact, in case of infection without vaccination, the curve of the infected
group goes to zero in about 90 days (see Figure~\ref{SIR_vac_simul1}), where in the
presence of vaccination the curve of the infected group goes to zero in about 45 days
(see Figure~\ref{SIR_vac_simul2}). This shows the efficiency of using vaccination
in controlling Ebola. Table~\ref{tab1} presents the maximum percentage
of infected individuals, corresponding to different rates of vaccinations,
and the respective number of days for the peak of infected.

\begin{table}
\centering
\begin{tabular}{|| c | c | c||}
\hline
\hline
\multirow{2}{*}{Rate of vaccination} & Maximum percentage & Days to the peak of\\ 
 & of infected & $I(t)$\\ \hline
$0$ & $17\%$  & $26.52$\\ \hline
$0.005$ & $15\%$  & $23.47$\\ \hline
$0.01$ & $13\%$  & $22.07$ \\ \hline
$0.02$ & $10\%$  & $17.09$ \\ \hline
$0.03$ & $9\%$  & $13.68$ \\ \hline
$0.06$ & $7\%$  & $9.28$ \\ \hline \hline
\end{tabular}
\caption{Maximum percentage of infected individuals $I$
corresponding to different rates $\upsilon$ of vaccination
and respective number of days for the peak of $I$.\label{tab1}}
\label{table:nonlin}
\end{table}


\section{The Optimal Control Problem}
\label{sec_control}

Optimal control has been recently used with success
in a series of epidemiological problems: see
\cite{delf,MR2719552,MR3266821,MR3101449}
and references therein. In this section,
we present an optimal control problem by introducing
into the model \eqref{SIR_model} a control $u(t)$ representing the vaccination rate
at time $t$. The control $u(t)$ is the fraction of susceptible
individuals being vaccinated per unit of time. We assume that all susceptible
individuals that are vaccinated are transferred directly to the removed class.
Then, the mathematical model with control is given by the following
system of nonlinear differential equations:
\begin{equation}
\label{SIR_control}
\begin{cases}
\dfrac{dS(t)}{dt} = -\beta S(t)I(t) - u(t) S(t),\\[0.3cm]
\dfrac{dI(t)}{dt} = \beta S(t)I(t) - \mu I(t),\\[0.3cm]
\dfrac{dR(t)}{dt} = \mu I(t) + u(t) S(t),
\end{cases}
\end{equation}
where $S(0) \geq 0$, $I(0) \geq 0$, and $R(0) \geq 0$ are given.
Our goal is to reduce the infected individuals and the cost of vaccination.
Precisely, our optimal control problem consists of minimizing the objective functional
\begin{equation}
\label{cost_func}
J(u) = \int_{0}^{t_{f}} I(t) + \dfrac{A}{2}u^2(t) dt,
\end{equation}
subject to $0\leq u(t)\leq 0.9$ for $t\in[0,t_{f}]$,
where the parameters $A\geq0$ and $t_f$ denote,
respectively, the weight on cost
and the duration of the vaccination program.
We present the numerical solution to this optimal control problem,
comparing the results with the numerical simulation results obtained
in Section~\ref{sec:vac} via \textsf{Matlab}. As in Section~\ref{sec:vac},
the rate of infection is given by $\beta=0.2$, the recovered rate is given by $\mu=0.1$,
and we use $S(0)=0.95$, $I(0)=0.05$, and $R(0)=0$ for
the initial number of susceptible, infected, and recovered populations, respectively
(i.e., at the beginning 95\% of population is susceptible and 5\% is already infected with Ebola).
Note that when $u(t) \equiv 0$ model \eqref{SIR_control} reduces to \eqref{SIR_model};
and when $u(t) \equiv \upsilon$ model \eqref{SIR_control} reduces to \eqref{SIR_vac}.

The numerical simulation of the system without control ($u(t) \equiv 0$)
was done with \textsf{Matlab} in Section~\ref{sec:Kep}.
For our numerical solution of the optimal control problem, we have used the
\textsf{ACADO} solver  \cite{acado}, which is based on a multiple shooting method,
including automatic differentiation and based ultimately on the semidirect 
multiple shooting algorithm of Bock and Pitt \cite{acado2}. \textsf{ACADO} is
a self-contained public domain software environment written
in \textsf{C++} for automatic control and dynamic optimization.
\begin{figure}
\centering
\includegraphics[width=10cm]{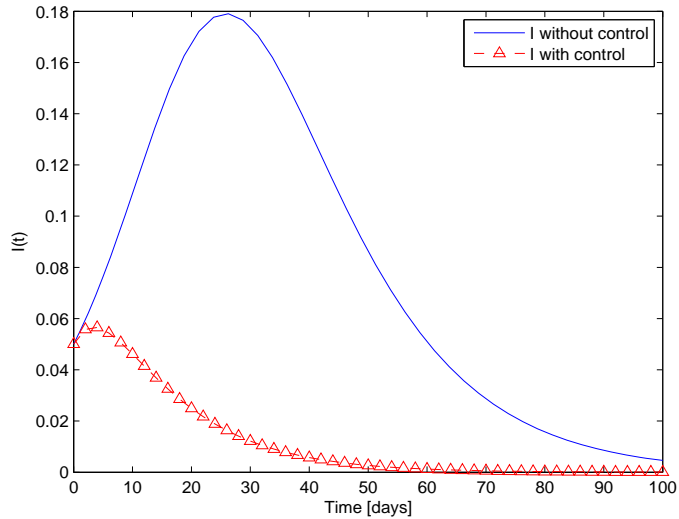}
\caption{The function $I(t)$ of infected individuals 
with and without control. \label{result1_I}}
\end{figure}
\begin{figure}
\centering
\includegraphics[width=10cm]{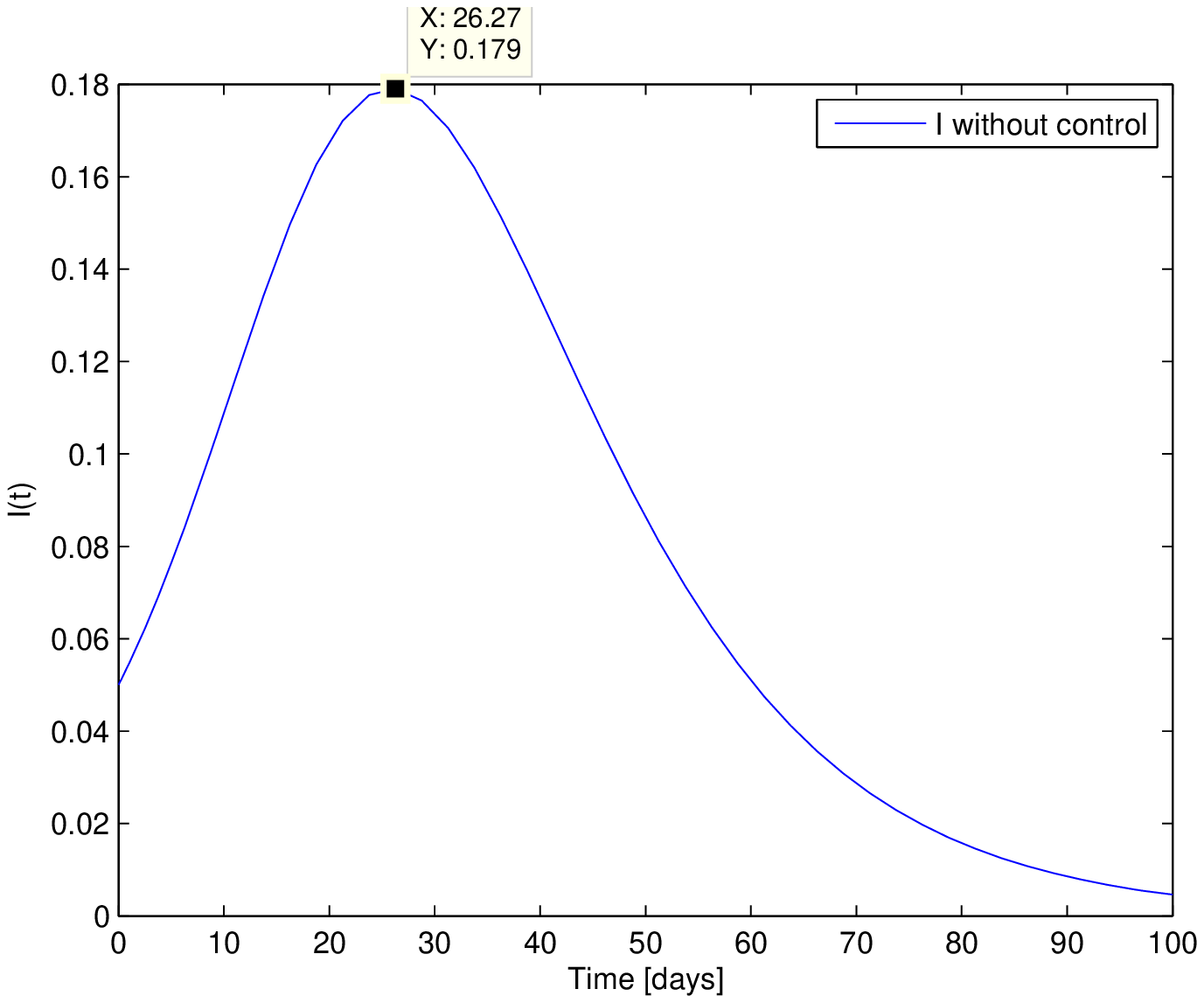}
\includegraphics[width=10cm]{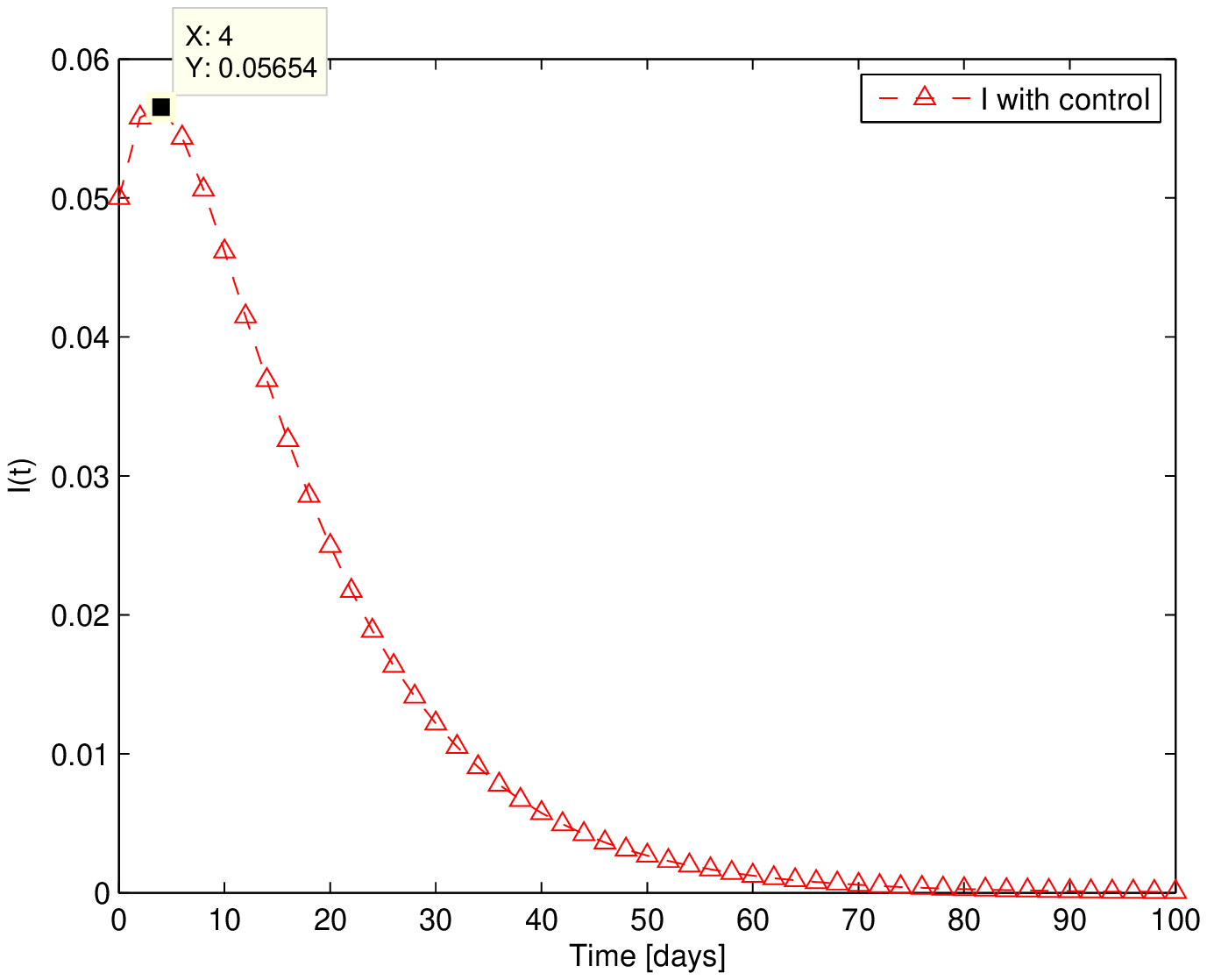}
\caption{Comparison between the peaks of infected individuals $I(t)$
in case of optimal control \emph{versus} without control. \label{result2_I}}
\end{figure}
\begin{figure}
\centering
\includegraphics[width=10cm]{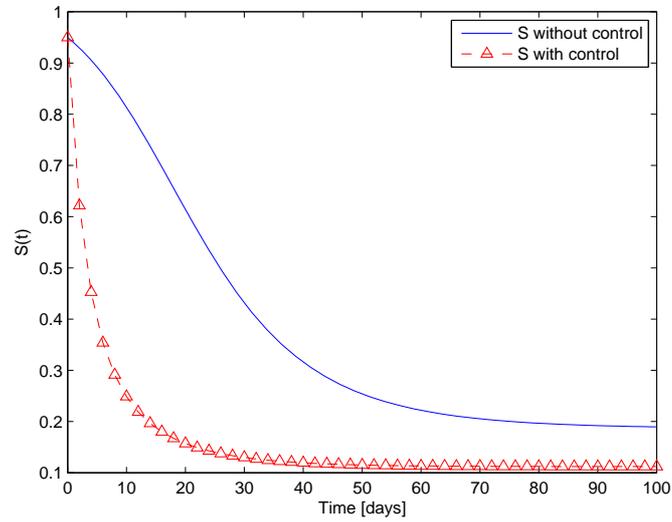}
\caption{Susceptible individuals $S(t)$
in case of optimal control \emph{versus} without control.\label{result_S}}
\end{figure}
\begin{figure}
\centering
\includegraphics[width=10cm]{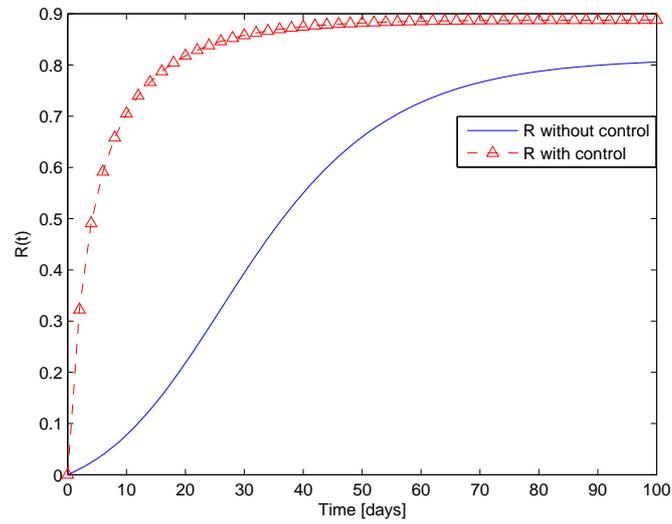}
\caption{Recovered individuals $R(t)$ in case of optimal control
\emph{versus} without control.\label{result_R}}
\end{figure}
\begin{figure}
\centering
\includegraphics[width=10cm]{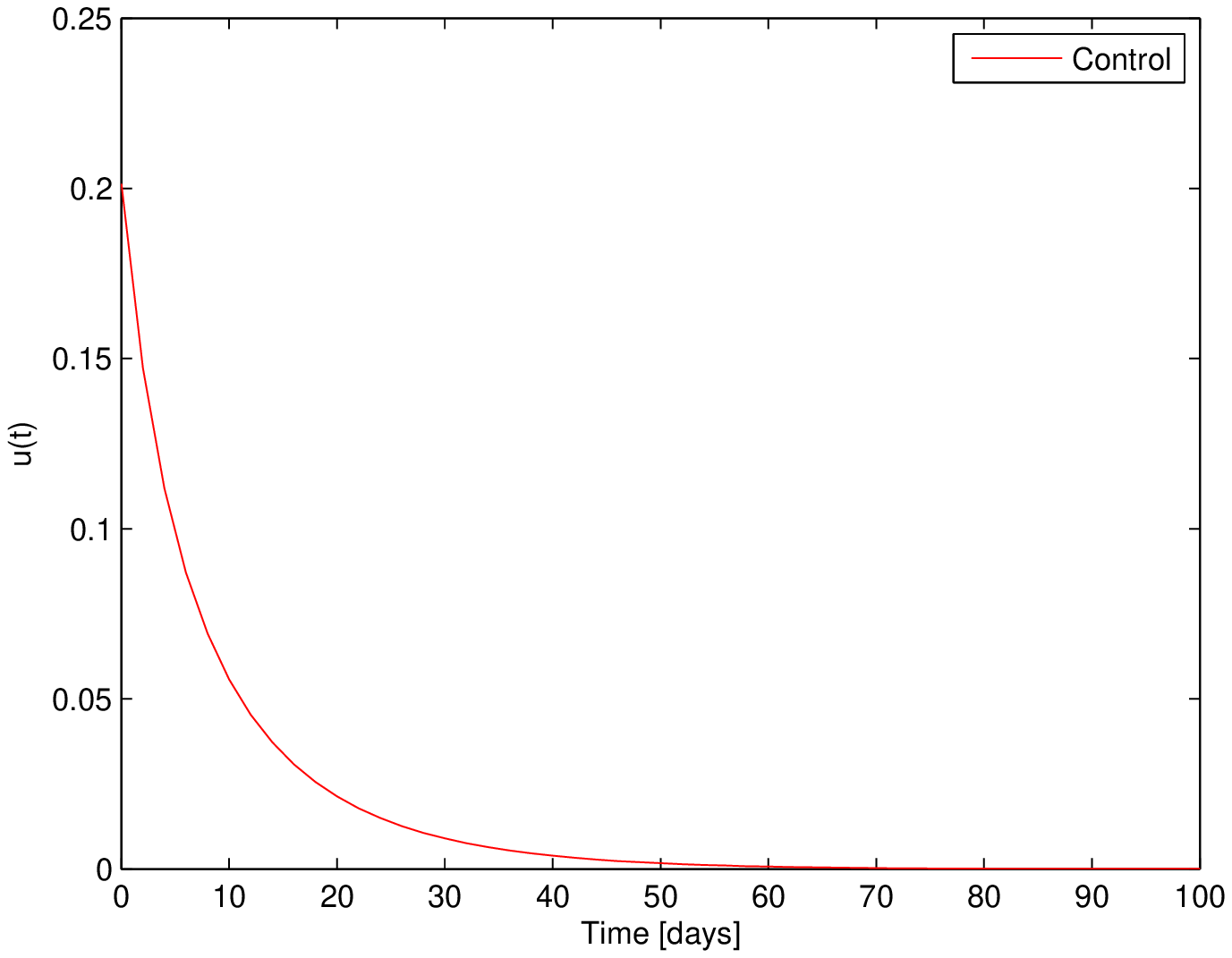}
\caption{The optimal control $u$ for problem \eqref{SIR_control}--\eqref{cost_func}.
\label{result_u}}
\end{figure}
Figures~\ref{result1_I} and \ref{result2_I} show the significant
difference in the number of infected individuals with and without control.
In fact, in case of optimal control, the function $I$ decreases greatly:
the maximum value on the infected curve $I$ under optimal control is 5.6\%
against 17.9\% without any control (see Figure~\ref{result2_I}).

In Figure~\ref{result_S}, we see that the number of susceptible $S$ decreases more rapidly
during the vaccination campaign. It reaches 11\% at the end of the campaign
against 19\% in the absence of optimal control.

Figure~\ref{result_R} shows that the number of recovered individuals increases rapidly.
The number $R(t_f)$ at the end of the optimal control vaccination period is 88.7\%
instead of 80.5\% without control. Figure~\ref{result_u} gives
a representation of the optimal control $u(t)$.


\section{Conclusion}
\label{sec:4}

We investigated a mathematical model that provides a good description of the 2014
Ebola outbreak in Liberia: the model fits well the data of confirmed cases in Liberia
provided by the World Health Organization. The evolution of infected individuals shows
that the infected population remains important, describing the actual evolution
of the virus in Liberia. This evolution is different than Ebola's outbreak in the last decades.
Our study of the SIR model with vaccination shows that vaccination
is a very efficient factor in reducing the number of infected individuals
in a short period of time and increasing the number of recovered individuals.
We also discussed an optimal control problem related to the impact
of a vaccination strategy on the spread of the 2014 outbreak of Ebola in West Africa.
The numerical simulation of both systems, with and without control,
shows that an optimal control strategy greatly helps to reduce the number of infected
and susceptible individuals and increases the number of recovered individuals.

As future work, we plan to include in our study other factors. In particular,
we intend to include in the mathematical model the treatment
of infected individuals with a quarantine procedure. Another interesting
line of research is to investigate the effect of impulsive vaccination.


\section*{Acknowledgments}

This research was initiated while Rachah was visiting the Department of Mathematics of
University of Aveiro, Portugal. The hospitality of the host
institution and the financial support provided by the
\emph{Institut de Math\'{e}matiques de Toulouse}, France,
are here gratefully acknowledged. Torres was supported by funds
through the Portuguese Foundation for Science and Technology (FCT),
within CIDMA project UID/MAT/04106/2013
and OCHERA project PTDC/EEI-AUT/1450/2012, cofinanced by FEDER
under POFC-QREN with COMPETE reference FCOMP-01-0124-FEDER-028894.



\end{document}